\documentclass[12pt]{article}
\usepackage{a4wide}
\usepackage{graphicx}           %   usepackage without driver option
\usepackage{color}
\newcommand{\addspan}[1]{{\color{black}#1}}
\newcommand{\addedSecond}[1]{{\color{black}#1}}
\newcommand{\cred}[1]{{\color{black}#1}}
\newcommand{\cmag}[1]{#1}
\newcommand{\proof}[1]{#1}
\newcommand{\err}[1]{{\color{black}#1}}

\usepackage{amsmath,amssymb}
\usepackage{epsf}

\newcommand{\GeV}{\,\text{GeV} }

\newcommand{\nn}{\nonumber\\}

\newcommand{\MSbar}{\ensuremath{\overline{\text{MS}}} }

\begin{document}

\title{\vbox{
\baselineskip 14pt
\hfill \hbox{\normalsize KUNS-2420}
} \vskip 1cm
\bf \Large Bare Higgs mass at Planck scale \vskip 0.5cm
}
\author{
Yuta~Hamada,\thanks{E-mail: \tt hamada@gauge.scphys.kyoto-u.ac.jp}~
Hikaru~Kawai,\thanks{E-mail: \tt hkawai@gauge.scphys.kyoto-u.ac.jp}~ 
and Kin-ya~Oda\thanks{E-mail: \tt odakin@gauge.scphys.kyoto-u.ac.jp}\bigskip\\
%\\*[20pt]
{\it \normalsize
Department of Physics, Kyoto University, Kyoto 606-8502, Japan}\smallskip
}
\date{\today}

%\newpage

%\begin{titlepage}
\maketitle
%\thispagestyle{empty}
%\clearpage
%\tableofcontents
%\thispagestyle{empty}
%\end{titlepage}

\abstract{\noindent \normalsize
%In a certain quantum gravity/string theory context, it is possible that not only the physical Higgs mass but also the bare one (and hence the radiative corrections as well) can vanish at the Planck/string scale.
We compute one- and two-loop quadratic divergent contributions to the bare Higgs mass in terms of the bare couplings in the Standard Model. %We identify the recently observed particle around 126\,GeV at the LHC as the SM Higgs boson and 
We approximate the bare couplings, defined at the ultraviolet cutoff scale, by the $\MSbar$ ones at the same scale, which are evaluated by the two-loop renormalization group equations for the Higgs mass around 126\,GeV in the Standard Model. 
We obtain the cutoff scale dependence of the bare Higgs mass, and examine where it becomes zero. %loop corrections to the Higgs mass is fine tuned to be zero. %when the bare Higgs mass is also vanishing. 
%We call this particular cutoff the fine-tuned scale. 
We find that when we take the current central value for the top quark pole mass, 173\,GeV, the bare Higgs mass vanishes if the cutoff is about $10^{23}$\,GeV. With a 1.3\,$\sigma$ smaller mass, 170\,GeV, the scale %of the fine-tuning 
can be of the order of the Planck scale.
}

\newpage

\normalsize

\section{Introduction}

The ATLAS~\proof{\cite{:2012gk}} and CMS~\proof{\cite{:2012gu}} experiments at the Large Hadron Collider (LHC) observed a particle  at the 5$\sigma$ confidence level (C.L.), which is consistent with the Standard Model (SM) Higgs boson with mass
\begin{align}
m_H
	&=	\begin{cases}
		126.0\pm0.4\pm0.4 \GeV,		&   \text{ATLAS~\cite{:2012gk}},\\
		125.3\pm0.4\pm0.5  \GeV,	&	\text{CMS~\cite{:2012gu}}.
		\end{cases}
	\label{observed Higgs mass}
\end{align}
Such a relatively light Higgs boson is compatible with the electroweak precision data~\cite{Beringer:1900zz}. Furthermore, this value of Higgs mass allows the SM to be valid up to the Planck scale, within the unitarity, (meta)stability, and triviality bounds~\cite{SM stability,Degrassi:2012ry,Alekhin:2012py}. Up to now, there are no symptoms of breakdown of the SM as an effective theory below the Planck scale.

On the other hand, if one wants to solve the Higgs mass fine-tuning problem within a framework of quantum field theory, it would be natural to assume a new physics at around the TeV scale. The supersymmetry is a possible solution to cancel the quadratic divergences in the Higgs mass, see e.g.\ Ref.~\cite{Martin:1997ns}. However, a Higgs mass around 126\GeV requires some amount of fine-tuning in the Higgs sector in the minimal supersymmetric Standard Model; see, e.g., Ref.~\cite{Baer:2012uy}. Furthermore, no sign of supersymmetry has been observed at LHC so far~\cite{nosusy}.

Given the current experimental situation, it is important to examine a \cred{possibility} in which the SM is valid towards a very high ultraviolet (UV) cutoff scale~$\Lambda$. \cred{In such a case, a fine-tuning of the Higgs mass must be done, as is the case for the cosmological constant. There are several approaches to the fine-tuning. One is simply not to regard it as a problem but to accept the parameters which nature has chosen. Instead, one may resort to the anthropic principle in which one explains the parameters by the necessity of the existence of ourselves; see, e.g., Refs.~\cite{Weinberg:1987dv,Hall:2009nd}. Or else, the tuning may be accounted for by quantum gravitational nonperturbative effects such as those from a multiverse or baby universe; see, e.g., Ref.~\cite{multiverse}. There are yet other discussions that the tuning is achieved within the context of field theory such as the classical conformal symmetry; see, e.g., Ref.~\cite{tuning within QFT}.
}

\cred{In this paper, we do not try to solve the naturalness problem. Rather, we evaluate the value of the bare parameters in order to investigate the Planck scale physics. They must be useful to connect the low energy physics to the underlying microscopic description, such as string theory.
}

In this paper, we compute the bare Higgs mass by taking into account one- and two-loop corrections in the SM. 
\cred{When we write in terms of the dimensionless bare couplings, the bare Higgs mass turns out to be a sum of a quadratically divergent part ($\propto\Lambda^2$), which is independent of the physical Higgs mass, and a logarithmically divergent one ($\propto \log\Lambda$). The importance of the coefficient of $\Lambda^2$ was first pointed out by Veltman at the one-loop order~\cite{Veltman:1980mj}. Generalizations to higher loops within the renormalized perturbation theory have been developed and applied in Ref.~\cite{others} in which the authors have reported the behavior $\sim\Lambda^2(\log\Lambda)^n$; see also Ref.~\cite{Drozd:2012is} for a review. In contrast, we see that such behavior does not appear in the bare perturbation theory. The reason why we employ the latter framework is that we are interested in the scale near the cutoff. These points will be discussed in detail with explicit calculations in Sec.~\ref{bare Higgs mass section}.}

We will see that the bare mass can be zero if $\Lambda$ is around the Planck scale, \cred{which gives some interesting suggestions on the Planck scale physics. First, it may imply that the supersymmetry of the underlying microscopic theory is restored above the Planck scale. In fact, superstring theory has many phenomenologically viable perturbative vacua in which supersymmetry is broken at the Planck scale; see, e.g., Ref.~\cite{Kawai:1986va}. In the last section, we will discuss that threshold corrections at the string scale may generate a small nonvanishing bare mass. Second, the vanishing of the bare Higgs mass together with that of the quartic Higgs coupling indicates almost flat potential near the Planck scale, which opens a possibility that the slow-roll inflation is achieved solely by the Higgs potential~\cite{ours}.
}

This paper is organized as follows. In the next section, we explain our convention, and calculate the quadratic divergent contributions to the bare Higgs mass up to the two-loop orders.
In Sec.~\ref{RGE_section}, we present a renormalization group equation (RGE) analysis in the SM and give our results for the Higgs quartic coupling at high scales. In Sec.~\ref{result section}, we examine how small the bare Higgs mass can be at the Planck scale and show at what scale the bare Higgs mass vanishes. We vary $\alpha_s$, $m_H$, and $m_t^\text{pole}$ to see how the results are affected. The last section contains the summary and discussions.

\newpage
\section{Bare Higgs mass}\label{bare Higgs mass section}
\addspan{In this section, we compute the quadratic divergence in the bare Higgs mass.}

\subsection{Bare mass in $\phi^4$ theory}\label{phi4 section}
\addedSecond{
Let us explain our treatment of the bare mass by taking a simple example of the $\phi^4$ theory with the bare Lagrangian:
\begin{equation}
\mathcal{L}=\frac{1}{2}(\partial_\mu \phi_B)^2-\frac{m_B^2}{2}\phi_B^2-\frac{\lambda_B}{4!}\phi_B^4.
	\label{phi4 bare Lagrangian}
\end{equation}
In the mass independent renormalization scheme,\footnote{
\addedSecond{See, e.g., the introduction and the subsequent section of Ref.~\cite{Fujikawa:2011zf} for a recent review of the discussion explained in this paragraph. In particular, our Eq.~\eqref{phi4 bare Lagrangian} corresponds to Eq.~(2.6) in Ref.~\cite{Fujikawa:2011zf}. Note that in Ref.~\cite{Fujikawa:2011zf} ``bare mass'' refers to $m_0$ whereas our terminology is the same as ``the common definition'', that is, we call $\Delta_\text{sub}+m_0^2$ the bare mass in general, though we consider only the leading term $\Delta_\text{sub}$ in actual computation.}
}
the bare mass $m_B^2$ is separated into the quadratically divergent part $\Delta_\text{sub}$ and the remaining one $m_0^2$:}
\begin{align}
\addedSecond{m_B^2=\Delta_\text{sub}+m_0^2.}
\end{align}
%One should not confuse the bare mass $m_B^2$ with the running mass \addedSecond{parameter} in a mass independent renormalization scheme~\cite{Weinberg:1951ss}.
\addedSecond{Here $\Delta_\text{sub}$ is chosen in such a way that the physical mass becomes zero when $m_0^2=0$.
Then the mass parameter \addedSecond{$m_0^2$} is introduced to describe the deviation from it and is multiplicatively renormalized to absorb the logarithmic divergence.} 
\addedSecond{We note that in the dimensional regularization, $\Delta_\text{sub}$ happens to be formally zero and only $m_0^2$ remains.\footnote{
\addedSecond{If one insists on the dimensional regularization, one might check the $D=2$ pole to see the quadratically divergent bare mass, which is beyond the scope of this paper.}
}
What we discuss in this paper is not \addedSecond{$m_0^2$} but the whole $m_B^2$. Since $m_0^2$ is negligibly small compared to $\Delta_\text{sub}$, we concentrate on the quadratically divergent part $\Delta_\text{sub}$ in the following.
}

%\addedSecond{As said above,} we explain our procedure by taking the concrete evaluation for the $\phi^4$ theory. 
From the bare Lagrangian~\eqref{phi4 bare Lagrangian}, we calculate the bare mass $m_B^2$ order by order in the loop expansion so that the physical mass is tuned to be zero\footnote{
Precisely speaking, $m^2_{B,\,\text{0-loop}}$ corresponds to the physical mass times the wave function renormalization factor and is negligibly small compared to the UV cutoff scale.
}
\begin{equation}
m_B^2=m_{B,\,\text{0-loop}}^2+ m_{B,\,\text{1-loop}}^2+ m_{B,\,\text{2-loop}}^2+\cdots.
\end{equation}
%のように$m_B^2$をloop展開し$\lambda$の各次数で$m_B^2$を決定する。
At each order, we fix the bare mass as
%\begin{fmffile}{phi4}
\begin{align}
m_{B,\,\text{0-loop}}^2
	&=	0,\\
m_{B,\,\text{1-loop}}^{2}
+i\left.\left(
          \parbox{25mm}{\includegraphics[width=25mm]{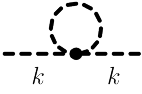}}
	\right)
      \right|_{k=0}  &=0, 
      \label{one-loop}
      \\
m_{B,\,\text{2-loop}}^{2}
%       \parbox{25mm}{\begin{fmfgraph*}(70,70)
%\fmfcmd{%
%   vardef cross_bar (expr p, len, ang)=
%   ((-len/2,0)--(len/2,0))
%      rotated (ang+angle direction length(p)/2 of p)
%      shifted point length(p)/2 of p
%      enddef;
%      style_def crossed expr p =
%         cdraw p;
%         ccutdraw cross_bar (p, 5mm,45);
%         ccutdraw cross_bar (p, 5mm, -45)
%       enddef;}
%              \fmfpen{thick}
%                \fmfleft{i1}
%                 \fmfright{o1}
%                 \fmf{dashes}{i1,v,o1}
%                   \fmf{crossed,label={$m_{B,\,\text{1-loop}}^2$}}{v,v}
%                   \fmfdot{v}
%             \end{fmfgraph*}}\ \ \ \ \
+i\left.\left(
          \parbox{25mm}{\includegraphics[width=25mm]{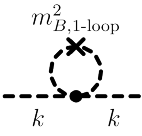}}
+
          \parbox{25mm}{\includegraphics[width=25mm]{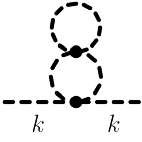}} %\ \ \ \ \
+
        \parbox{25mm}{\includegraphics[width=25mm]{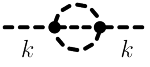}}
                      \label{two-loop}
            % \ \ \ \ \
\right)\right|_{k=0}
	&=	0.
\end{align}
The one-loop integral in Eq.~\eqref{one-loop} is quadratically divergent and is proportional to
\begin{align}
	I_1	&:=	\int \frac{d^4p_E}{(2\pi)^4}\frac{1}{p_E^2},
		\label{I1}
\end{align}
where $p_E$ is a Euclidean four momentum. In the two-loop computation~\eqref{two-loop}, the momentum integrals in the third and fourth terms are, respectively,
\begin{align}
J_2	&:=	\int \frac{d^4p_E}{(2\pi)^4}\frac{d^4 q_E}{(2\pi)^4}\frac{1}{p_E^4 q_E^2},
		\label{J2}\\
I_2	&:=	\int \frac{d^4p_E}{(2\pi)^4}\frac{d^4 q_E}{(2\pi)^4}\frac{1}{p_E^2 q_E^2 (p_E+q_E)^2}.
		\label{I2}
\end{align}
The integral $J_2$ is infrared (IR) divergent: $J_2\propto \Lambda^2\ln(\Lambda/\mu_\text{IR})$, but is canceled by the second term in Eq.~\eqref{two-loop} due to the lower order condition~\eqref{one-loop}. Therefore, we are left with only $I_2$, which does not suffer from the infrared divergence. This situation does not change in higher orders because a mass should not contain an IR divergence.

\subsection{Bare mass in SM}

For the SM Higgs sector, we start from the bare Lagrangian of the following form in a fixed cutoff scheme with cutoff $\Lambda$:\footnote{\label{no higher dimensional}
\cred{In general the effective Lagrangian of an underlying microscopic theory at the cutoff scale contains higher dimensional operators. Their effects can be absorbed by the redefinition of the renormalizable and super-renormalizable couplings in the low energy region.
Therefore it suffices to take the form of Eq.~\eqref{SM bare Lagrangian} without higher dimensional operators in order to reproduce the low energy physics. 
However, the differences among the bare theories emerge when the energy scale gets close to the cutoff $\Lambda$.}
}
\begin{align}
\mathcal{L}
	&=	(D_\mu \phi_B)^\dagger(D^\mu \phi_B)-m_B^2\phi_B^\dagger \phi_B -\lambda_B(\phi_B^\dagger \phi_B)^2,	&
\phi_B
	&=
\begin{pmatrix}
\phi_B^+ \\
\phi_B^0
\end{pmatrix}.
	\label{SM bare Lagrangian}
\end{align}
%Hereafter throughout this paper, we omit the subscript $B$ for the bare Higgs mass and write $m_B^2$. In this section, we also omit $B$ from the coupling constants since they are all bare.
\cred{We set the physical mass to be zero: $m_\text{$B$,0-loop}^2=0$, as we are interested in physics at very high scales.}\footnote{\cred{
We are not intending to realize the Coleman-Weinberg mechanism, but to neglect the physical mass that is unimportant for our consideration.}}
The Planck scale is
\begin{equation}
M_{\text{Pl}}=\frac{1}{\sqrt{G_N}}=1.22\times10^{19} \GeV.
\end{equation}
We take into account the SM couplings $g_Y$, $g_2$, $g_3$, $\lambda$, $y_t$ and neglect the others. 
%A physical pole mass of a particle is always written by a capital $M$.

Now let us \addedSecond{follow} the prescription\addedSecond{, shown in the previous subsection,} in the SM. In the following, we work in the symmetric phase $\left\langle\phi\right\rangle=0$ as we are interested only in the quadratic divergent terms. In the evaluation of the Feynman diagrams, it is convenient to take the Landau gauge  for all the $SU(3)\times SU(2)\times U(1)$ gauge fields. In this gauge, a diagram always vanishes if an external Higgs line is attached with a gauge boson propagator by a three-point vertex:
\begin{align}
\left.
        \parbox{25mm}{\includegraphics[width=25mm]{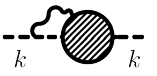}}
\right|_{k=0}
	&=	0.
\end{align}
From the one-loop diagrams we get the quadratic divergent integral $I_1$ again~\cite{Veltman:1980mj}
\begin{align}
m_{B,\,\text{1-loop}}^2
	&=	-\bigg(6\lambda_B+\frac{3}{4} g_{YB}^2+\frac{9}{4} g_{2B}^2-6y_{tB}^2\bigg)\,I_1.
	\label{Veltman 1-loop}
\end{align}

\begin{figure}[t]
\begin{center}
\includegraphics[width=\textwidth]{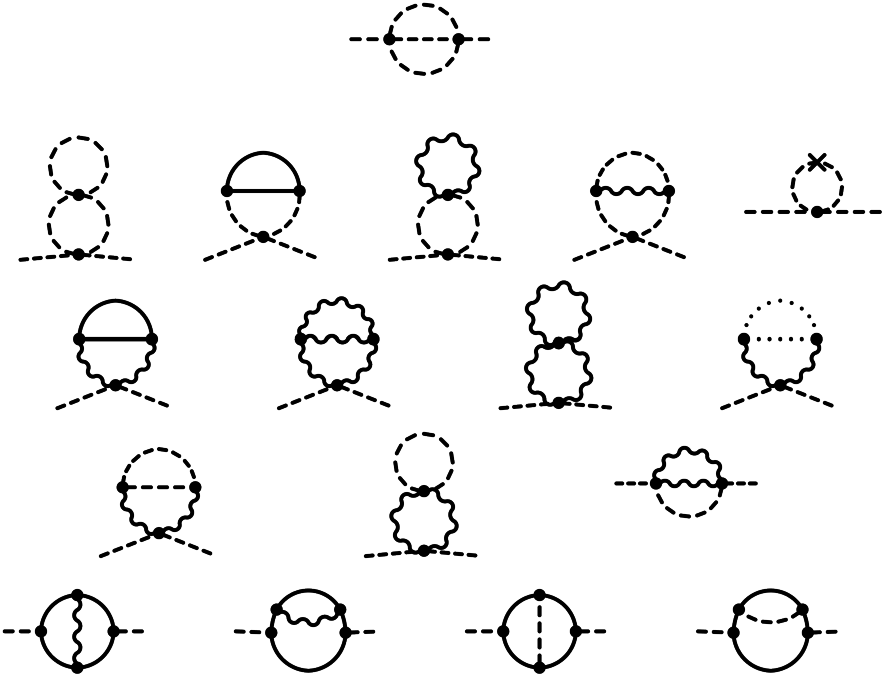}
\caption{Nonvanishing two-loop Feynman diagrams. Arrows are omitted. The dashed, solid, wavy, and dotted lines represent the scalar, fermion, gauge, and ghost propagators, respectively.}\label{two-loop diagrams}
\end{center}
\end{figure}
%\end{fmffile}
In Fig.~\ref{two-loop diagrams}, we present the two-loop Feynman diagrams that do not vanish  in the symmetric phase $\left\langle\phi\right\rangle=0$ and in the Landau gauge.
In the second row of Fig.~\ref{two-loop diagrams}, the last diagram cancels the divergences coming from the one-loop self-energy of the internal Higgs propagators, as in Eq.~\eqref{two-loop}.\footnote{
In practice, from each diagram containing a self-energy correction, one subtracts a term that is obtained by setting the external momentum of its self-energy to zero. \addedSecond{We have also applied} this subtraction for diagrams containing a vacuum polarization.
\addedSecond{For the gauge boson, this subtraction introduces a bare mass, which becomes zero in a gauge invariant regularization scheme such as the Pauli-Villars or dimensional regularizations.}
}
All the momentum integrals can be recast into either $I_2$ or $J_2$.\footnote{
\addedSecond{
Gauge invariance is formally satisfied in the sense that the Ward-Takahashi identity holds if we shift momenta freely without worrying about the ultraviolet divergences. In this paper, we are interested in the quadratic divergences that are left after these momentum redefinitions.}
}
We have explicitly checked that the coefficients of the infrared divergent integral $J_2$ cancel in each gauge invariant set of diagrams.\footnote{
\err{More precisely, we have assumed existence of a gauge invariant regularization behind, and have subtracted the quadratic divergences in the one-loop vacuum polarization. The cancellation of the coefficients of $J_2$ is checked under this assumption.}
}
\err{We then obtain the $g^4$ terms in $m_{B,\,\text{2-loop}}^2$ as in Table~\ref{detail}.}
\begin{table}
\begin{center}
\caption{$g^4$ terms in $m_{B,\,\text{2-loop}}^2$ in units of $I_2$.\label{detail}}
\includegraphics[width=0.6\textwidth]{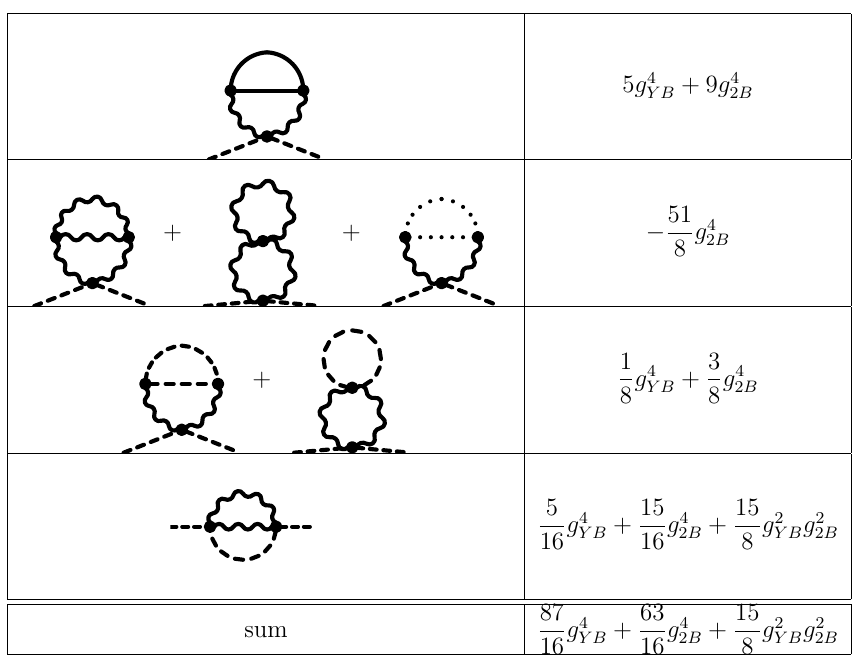}
\end{center}
\end{table}

By collecting these terms,
the two-loop contribution to the bare Higgs mass at $\Lambda$ becomes\footnote{As mentioned in Ref.~\cite{Veltman:1980mj}, while at the one-loop level, only a restricted set of particles participates; on the two-loop level, all kinds of particles up to the Planck mass enter in the discussion. We assume that there appear only SM degrees of freedom up to the UV cutoff scale.}
\begin{align}
m_{B,\,\text{2-loop}}^2
	&=	-\bigg\{
			9y_{tB}^4
			+y_{tB}^2\left(-\frac{7}{12}g_{YB}^2+\frac{9}{4}g_{2B}^2-16g_{3B}^2\right)\nonumber \\
	&\phantom{=	-\bigg\{}
			\err{
				\mbox{}-\frac{87}{16} g_{YB}^4-\frac{63}{16} g_{2B}^4-\frac{15}{8} g_{YB}^2g_{2B}^2
				}
			\nn
	&\phantom{=	-\bigg\{}
		+\lambda_B\left(-18y_{tB}^2+3 g_{YB}^2+9 g_{2B}^2\right)
		-\err{12}\lambda_B^2
		\bigg\}\,I_2.
		\label{bare mass in SM}
\end{align}
This is one of our main results.
Note that Eqs.~\eqref{Veltman 1-loop} and \eqref{bare mass in SM} are minus the radiative corrections to the physical Higgs mass squared; see Eqs.~\eqref{one-loop} and \eqref{two-loop}.

In Sec.~\ref{result section}, we will examine whether the bare mass can vanish at a particular UV cutoff scale. For that purpose, we need to relate the integrals $I_1$ and $I_2$. This relation necessarily depends on the cutoff scheme.\footnote{
\addedSecond{One can rigorously compute both $I_1$ and $I_2$ in principle if one fixes a cutoff scheme, such as an embedding in string theory. For our purpose, the simplified procedure~\eqref{our regularization} suffices as we just want to check the smallness of the two-loop contributions.}
}
In particular, if the two-loop contribution to the bare mass $m^2_{B,\,\text{2-loop}}$ becomes sizable compared to $m^2_{B,\,\text{1-loop}}$, the result suffers from a large theoretical uncertainty. We will verify that it is actually small. With this caution in mind, let us employ the following regularization:
\begin{equation}
\int d^4k_E\frac{1}{k_E^2}=\int_\varepsilon^\infty d\alpha \int d^4k_E\,e^{-\alpha k_E^2},
	\label{our regularization}
\end{equation}
which gives
\begin{align}
I_1&=\frac{1}{\varepsilon}\frac{1}{16\pi^2}, &
I_2&=\frac{1}{\varepsilon}\frac{1}{(16\pi^2)^2}\ln{\frac{2^6}{3^3}}\simeq0.005\,I_1.
	\label{two-loop in our scheme}
\end{align}
When we employ a naive momentum cutoff by $\Lambda$, we get
\begin{equation}
I_1=\frac{\Lambda^2}{16\pi^2},
\end{equation}
and hence we can regard $1/\varepsilon = \Lambda^2$.

%\addspan{We emphasize again that the regularization~\eqref{our regularization} is introduced just to verify that the ratio $I_2/I_1$ is small so that the two-loop contribution is indeed suppressed by the loop factor, regardless of the $O(1)$ ambiguity that is cutoff-scheme dependent. This fact will be used in the argument below Eq.~\eqref{bare mass at Planck scale}.}

\subsection{Graviton effects}\label{graviton section}
Let us estimate the graviton loop effects on the above obtained result. The graviton $h_{\mu\nu}$ in the metric
\begin{align}
g_{\mu \nu}&=\eta_{\mu \nu}+\frac{\sqrt{32\pi}}{M_{\text{Pl}}}h_{\mu \nu}
\end{align}
couples to the Higgs through the energy-momentum tensor:
\begin{align}
T_{\mu \nu}
	&=	\frac{2}{\sqrt{-g}}\frac{\delta}{\delta g^{\mu \nu}}\sqrt{-g}\mathcal{L}\nn
	&=	(D_\mu \phi)^\dagger(D_\nu \phi)+(D_\nu \phi)^\dagger(D_\mu \phi)
		-g_{\mu \nu}\left[
			(D_\mu \phi)^\dagger(D^\mu \phi)-m_B^2\phi^\dagger \phi -\lambda(\phi^\dagger \phi)^2
			\right].
\end{align}
The most divergent contributions come from two derivative couplings. A one-loop diagram containing such a graviton coupling vanishes because it necessarily picks up an external momentum, which is set to zero. Other contributions are at most logarithmically divergent. At the two-loop level, diagrams involving an internal graviton line that does not touch a Higgs external line give a form $\Lambda^4/M_\text{Pl}^2$. If the UV cutoff is much smaller than the Planck scale, this becomes negligible, and the higher loops become further insignificant. Indeed in perturbative string theory, higher loop corrections are proportional to powers of the string coupling constant $g_s$ and become subleading. If the cutoff scale exceeds the Planck scale, we cannot neglect the graviton contributions.

\section{SM RGE evolution toward Planck scale}\label{RGE_section}
In Sec.~\ref{result section}, we will approximate the \addspan{dimensionless} bare coupling constants in the SM at the UV cutoff scale~$\Lambda$ by the running ones in the modified minimal subtraction~($\MSbar$) scheme at the same scale $\Lambda$; see the Appendix %~\ref{cutoff vs MSbar} 
for its justification.
\addspan{We note that the \MSbar couplings will be used solely to approximate the dimensionless bare couplings at the cutoff scale and that the bare Higgs mass does not run.}

%\subsection{Two loop RGE in SM}
To get the $\MSbar$ running coupling constant, we apply the RGE at the two-loop order. For $g_Y$, $g_2$, $g_3$, and $y_t$, we use the ones in Ref.~\cite{Arason:1991ic}.\footnote{We replace $g_1$ of the GUT normalization to $g_Y=\sqrt{3/5}\,g_1$ and rewrite the quartic coupling as $\lambda_\text{\cite{Arason:1991ic}}=2\lambda$, where $\lambda_\text{\cite{Arason:1991ic}}$ is the one employed in Ref.~\cite{Arason:1991ic}.}
For the quartic coupling, we employ the one given in Ref.~\cite{Ford:1992pn}.\footnote{
We use the arXiv version 2 of Ref.~\cite{Ford:1992pn} with the replacements $g'=g_Y$, $g=g_2$, $h=y_t$, and $\lambda_\text{\cite{Ford:1992pn}}=6\lambda$, where $\lambda_\text{\cite{Ford:1992pn}}$ is the quartic coupling employed in Ref.~\cite{Ford:1992pn}. The RGE for $\lambda$ in Ref.~\cite{Arason:1991ic} becomes equal to that of Ref.~\cite{Ford:1992pn}, after correcting $-{3\over2}g_2^4Y_4(S)$ to $-{3\over2}g_2^4Y_2(S)$ and changing the part ${229\over4}+{50\over9}n_g$ to $\cmag{-}{229\over24}\cmag{-}{50\over9}n_g$ in Eq.~(A.17) in Ref.~\cite{Arason:1991ic}.}
To be explicit,
\begin{align}
\frac{dg_Y}{dt}
	&=	{1\over16\pi^2}\frac{41}{6}g_Y^3+\frac{g_Y^3}{(16\pi^2)^2}\left({199\over18}g_Y^2+{9\over2}g_2^2+{44\over3}g_3^2-{17\over6}y_t^2\right),\nn
\frac{dg_2}{dt}&=-{1\over16\pi^2}\frac{19}{6}g_2^3
	+\frac{g_2^3}{(16\pi^2)^2}\left({3\over2}g_Y^2+{35\over6}g_2^2+12g_3^2-{3\over2}y_t^2\right),\nn
\frac{dg_3}{dt}
	&=	-\frac{7}{16\pi^2}g_3^3+\frac{g_3^3}{(16\pi^2)^2}\left({11\over6}g_Y^2+{9\over2}g_2^2-26g_3^2-2y_t^2\right),\nn
\frac{dy_t}{dt}
	&=	\frac{y_t}{16\pi^2}\bigg(\frac{9}{2}y_t^2-\frac{17}{12}g_Y^2-\frac{9}{4}g_2^2-8g_3^2\bigg)+\frac{y_t}{(16\pi^2)^2}\bigg(-12y_t^4+6\lambda^2-12\lambda y_t^2\nn
	&\quad
		+\frac{131}{16}g_Y^2 y_t^2+\frac{225}{16}g_2^2 y_t^2+36 g_3^2 y_t^2+\frac{1187}{216}g_Y^4-\frac{23}{4}g_2^4-108g_3^4-\frac{3}{4}g_Y^2 g_2^2+9g_2^2 g_3^2+\frac{19}{9}g_3^2 g_Y^2\bigg),\nn
\frac{d\lambda}{dt}
	&=	\frac{1}{16\pi^2}\bigg(24 \lambda^2-3g_Y^2 \lambda-9g_2^2 \lambda+\frac{3}{8}g_Y^4+\frac{3}{4}g_Y^2 g_2^2 +\frac{9}{8}g_2^4+12\lambda y_t^2-6y_t^4\bigg)\nn
	&\quad
		+\frac{1}{(16\pi^2)^2}\bigg\{
			-312\lambda^3+36\lambda^2(g_Y^2+3g_2^2)
			-\lambda\left(\cmag{-}{629\over24}g_Y^4-{39\over4}g_Y^2g_2^2+{73\over8}g_2^4\right)\nn
	&\phantom{\quad+\frac{1}{(16\pi^2)^2}\bigg\{}
		+\frac{305}{16}g_2^6-\frac{289}{48}g_Y^2 g_2^4 -\frac{559}{48}g_Y^4 g_2^2 -\frac{379}{48}g_Y^6 -32 g_3^2 y_t^4-\frac{8}{3}g_Y^2 y_t^4-\frac{9}{4}g_2^4 y_t^2\nn
	&\phantom{\quad+\frac{1}{(16\pi^2)^2}\bigg\{}
		+\lambda y_t^2 \bigg(\frac{85}{6}g_Y^2+\frac{45}{2}g_2^2+80g_3^2\bigg)+g_Y^2 y_t^2\bigg(-\frac{19}{4}g_Y^2+\frac{21}{2}g_2^2\bigg)\nn
	&\phantom{\quad+\frac{1}{(16\pi^2)^2}\bigg\{}
		-144 \lambda^2 y_t^2-3\lambda y_t^4+30y_t^6\bigg\},
		\label{our RGE}
\end{align}
where $t=\ln\mu$. Though we do not include the bottom and tau Yukawa couplings in this paper, we have checked that these are negligible within the precision that we work in.

We put the boundary condition for the RGE~\eqref{our RGE} according to Ref.~\cite{Degrassi:2012ry}.
The $\MSbar$ gauge coupling of $SU(3)$ is given by the three-loop RGE running from $m_Z$ to $m_t^\text{pole}$ and matching with six flavor theory as~\cite{Degrassi:2012ry}
\begin{equation}
g_s(m_t^\text{pole})=1.1645+0.0031\bigg(\frac{\alpha_s(m_Z)-0.1184}{0.0007}\bigg)-0.00046\bigg(\frac{m_t^\text{pole}}{\GeV}-173.15\bigg),
\end{equation}
where $m_t^\text{pole}$ is the pole mass of the top quark. The $\MSbar$ quartic coupling at the top pole mass $m_t^\text{pole}$ is given by taking into account the QCD and Yukawa two-loop corrections~\cite{Degrassi:2012ry}
\begin{equation}
\lambda(m_t^\text{pole})=0.12577+0.00205\bigg(\frac{m_H}{\GeV}-125\bigg)-0.00004\bigg(\frac{m_t^\text{pole}}{\GeV}-173.15\bigg)\pm0.00140_\text{th},
\end{equation}
where $m_H$ is the observed Higgs mass which we read off from Eq.~\eqref{observed Higgs mass} as
\begin{align}
m_H	&=	125.7\pm0.6\GeV.
\end{align}
The $\MSbar$ top Yukawa coupling at the scale $m_t^\text{pole}$ is given by taking into account the QCD three-loop, electroweak one-loop, and $O(\alpha\alpha_s)$ two-loop corrections~\cite{Degrassi:2012ry}:
\begin{align}
y_t(m_t^\text{pole})
	&=	0.93587+0.00557 \bigg(\frac{m_t^\text{pole}}{\GeV}-173.15\bigg)-0.00003\bigg(\frac{m_H}{\GeV}-125\bigg)\nn
    &\quad
    	-0.00041 \bigg(\frac{\alpha_s(m_Z)-0.1184}{0.0007}\bigg)\pm0.00200_\text{th}.
\end{align}

In a more recent work~\cite{Alekhin:2012py}, it has been pointed out that the error in the top quark pole mass, consistently derived from the running one, is larger than that given in Ref.~\cite{Degrassi:2012ry}, $173.1\pm0.7\GeV$. The value obtained is~\cite{Alekhin:2012py}
\begin{align}
m_t^\text{pole}	&=	173.3\pm2.8\GeV,
\end{align}
which we will use in our analysis.

\begin{figure}[t]
%\begin{minipage}{0.33\hsize}
\begin{center}
	\hfill
  \includegraphics[width=.4\textwidth]{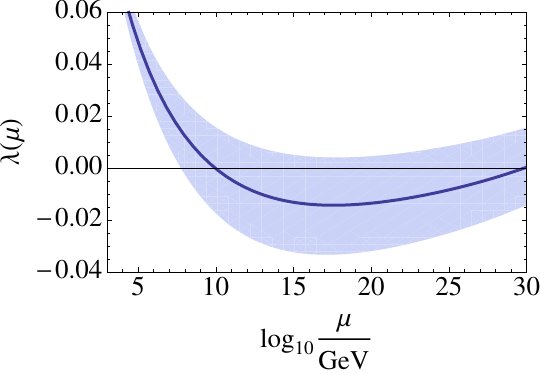}
	\hfill
  \includegraphics[width=.44\textwidth]{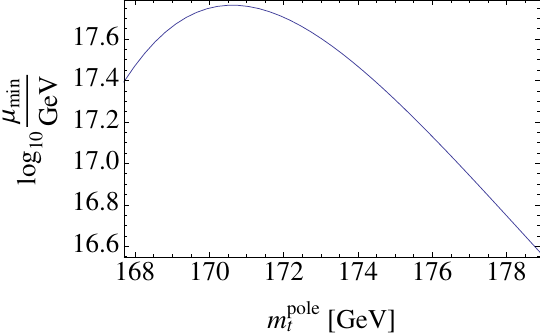}
	\hfill\mbox{}
  \caption{Left: $\MSbar$ running of the quartic coupling $\lambda$. The band corresponds to the $1\sigma$ deviation $m_t^\text{pole}=173.3\pm2.8\GeV$. Right: The scale $\mu_\text{min}$ at which $\lambda(\mu)$ takes its minimum value, as a function of $m_t^\text{pole}$. In both panels, low energy inputs are given by the central values $\alpha_s(m_Z)=0.1184$ and $m_H=125.7\GeV$.}
  \label{fig:lambda}
%\end{minipage}
\end{center}
\end{figure}
We plot the $\MSbar$ running coupling constant $\lambda(\mu)$ in Fig.~\ref{fig:lambda}.
As we increase the scale $\mu$,  the coupling $\lambda$ first decreases due to the term $-6y_t^4$ and remains small above $\mu=10^{10}\GeV$ for a while.
At further higher energies, $y_t$ becomes smaller and $\lambda$ starts to increase due to the contribution from $\frac{3}{8}g_Y^4$ which is not asymptotically free. At the intermediate scale, $\lambda$ can become negative but it is shown that a metastability condition can be met even in this case~\cite{SM stability,Degrassi:2012ry,Alekhin:2012py}.\footnote{\label{higher order terms}
\cred{At first sight, $\lambda_B<0$ seems to indicate a runaway potential. In the SM, radiative corrections from the top quark loop generates a potential barrier. The metastability argument does not assume an existence of a true stable vacuum at a very high scale but computes the vacuum decay rate from the area of the potential barrier from $\phi=0$ to the other zero point. In our case, it is possible that the runaway potential can be cured for a negative but small coupling ($\lambda_B<0, \left|\lambda_B\right|\ll1$) by the higher dimensional operators with positive couplings, such as $\left|\phi\right|^6/\Lambda^2$, which become important near the cutoff scale $\Lambda$. See also footnote~\ref{no higher dimensional}.}
}
The value of $\lambda$ at the Planck scale $M_{\text{Pl}}$ becomes consistent with Eq.~(64) in Ref.~\cite{Degrassi:2012ry}:
%\begin{align}
%\lambda(M_\text{Pl})
%	&=	-0.015
%		+0.0028\left({m_H\over\text{GeV}}-125\right)
%		-0.0046\left({m_t^\text{pole}-173.15\GeV\over0.7\GeV}\right)\nn
%	&\quad
%		+0.0018\left({\alpha_s(m_Z)-0.1184\over0.0007}\right)
%		\pm0.002_\text{th,$y_t(m_t^\text{pole})$}
%		\pm0.002_\text{th,$\lambda(m_t^\text{pole})$}.
%\end{align}
%\begin{align}
%\lambda(M_\text{Pl})
%	&=	-0.014
%		-0.018\left({m_t^\text{pole}-173.3\GeV\over2.8\GeV}\right)
%		+0.0018\left({\alpha_s(m_Z)-0.1184\over0.0007}\right)\nn
%	&\quad
%		+0.0017\left({m_H-125.7\GeV\over0.6\GeV}\right)
%		\pm0.004_\text{th}.
%\end{align}
\begin{align}
\lambda(M_\text{Pl})
	&=	-0.014
		-0.018\left({m_t^\text{pole}-173.3\GeV\over2.8\GeV}\right)
		+0.002\left({\alpha_s(m_Z)-0.1184\over0.0007}\right)\nn
	&\quad
		+0.002\left({m_H-125.7\GeV\over0.6\GeV}\right)
		\pm0.004_\text{th}.
		\label{lambda at UV cutoff}
\end{align}
As we can see from the left panel in Fig.~\ref{fig:lambda}, the value of the quartic coupling stays around its minimum in $10^{15}\GeV\lesssim \mu\lesssim 10^{20}\GeV$. Therefore, the minimum value of $\lambda$ is also given by Eq.~\eqref{lambda at UV cutoff} within our precision. In the right panel in Fig.~\ref{fig:lambda}, we plot $\mu_\text{min}$ at which the $\lambda(\mu)$ takes its minimum value. The central value $m_t^\text{pole}=173.3\GeV$ gives $\mu_\text{min}=4\times10^{17}\GeV$. 

%We give an explicit formula for the scale where $\lambda$ takes its minimum value and its value at the scale:
%\begin{align}
%\lambda_{\text{min}}
%	&=	-0.014
%		-0.018\left({m_t^\text{pole}-173.3\GeV\over2.8\GeV}\right)\nn
%%		+??\bigg(\frac{m_H}{\GeV}-125\bigg)+??\bigg(\frac{m_t^\text{pole}}{\GeV}-173.15\bigg)+??\bigg (\frac{\alpha_s(m_Z)-0.1184}{0.0007}\bigg)\pm??_\text{th}
%\log_{10}{\mu\over\text{GeV}}
%	&=	17.6
%		-0.4\left({m_t^\text{pole}-173.3\GeV\over2.8\GeV}\right)
%		-0.1\left({m_t^\text{pole}-173.3\GeV\over2.8\GeV}\right)^2
%		+??\bigg(\frac{m_t^\text{pole}}{\GeV}-173.15\bigg)+??\bigg (\frac{\alpha_s(m_Z)-0.1184}{0.0007}\bigg)\pm??_\text{th}
%\end{align}

%%%
\section{Bare Higgs mass at Planck scale}\label{result section}
%%%

%\subsection{Numerical results}
\begin{figure}[t]
%\begin{minipage}{0.5\hsize}
  \begin{center}
	\hfill
	\includegraphics[width=.4\textwidth]{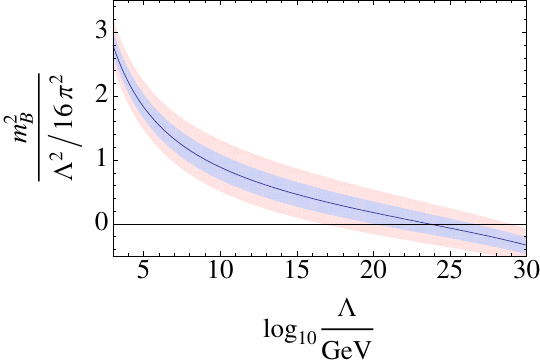}
	\hfill
	\includegraphics[width=.42\textwidth]{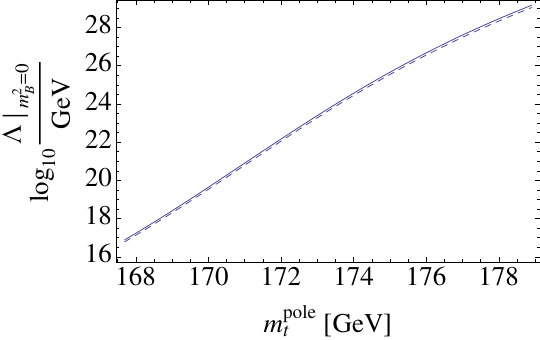}
	\hfill
	\mbox{}
	\end{center}
  \caption{Left: The bare Higgs mass $m_B^2$ in units of $\Lambda^2/16\pi^2$ vs the UV cutoff scale $\Lambda$. The blue (narrower) and pink (wider) bands represent the one and two sigma deviations of $m_t^\text{pole}$, respectively. Right: The UV cutoff scale at which the bare mass~$m_B^2$ becomes zero as a function of $m_t^\text{pole}$. The solid (dashed) line corresponds to the scale where $m_B^2$ ($m^2_{B,\,\text{1-loop}}$) becomes zero.  In both panels, we have taken the central values $\alpha_s(m_Z)=0.1184$ and $m_H=125.7\GeV$.}
  \label{fig:msq}
%\end{minipage}
\end{figure}
\begin{figure}[t]
%\begin{minipage}{0.5\hsize}
  \begin{center}
  \hfill
  \includegraphics[width=.5\textwidth]{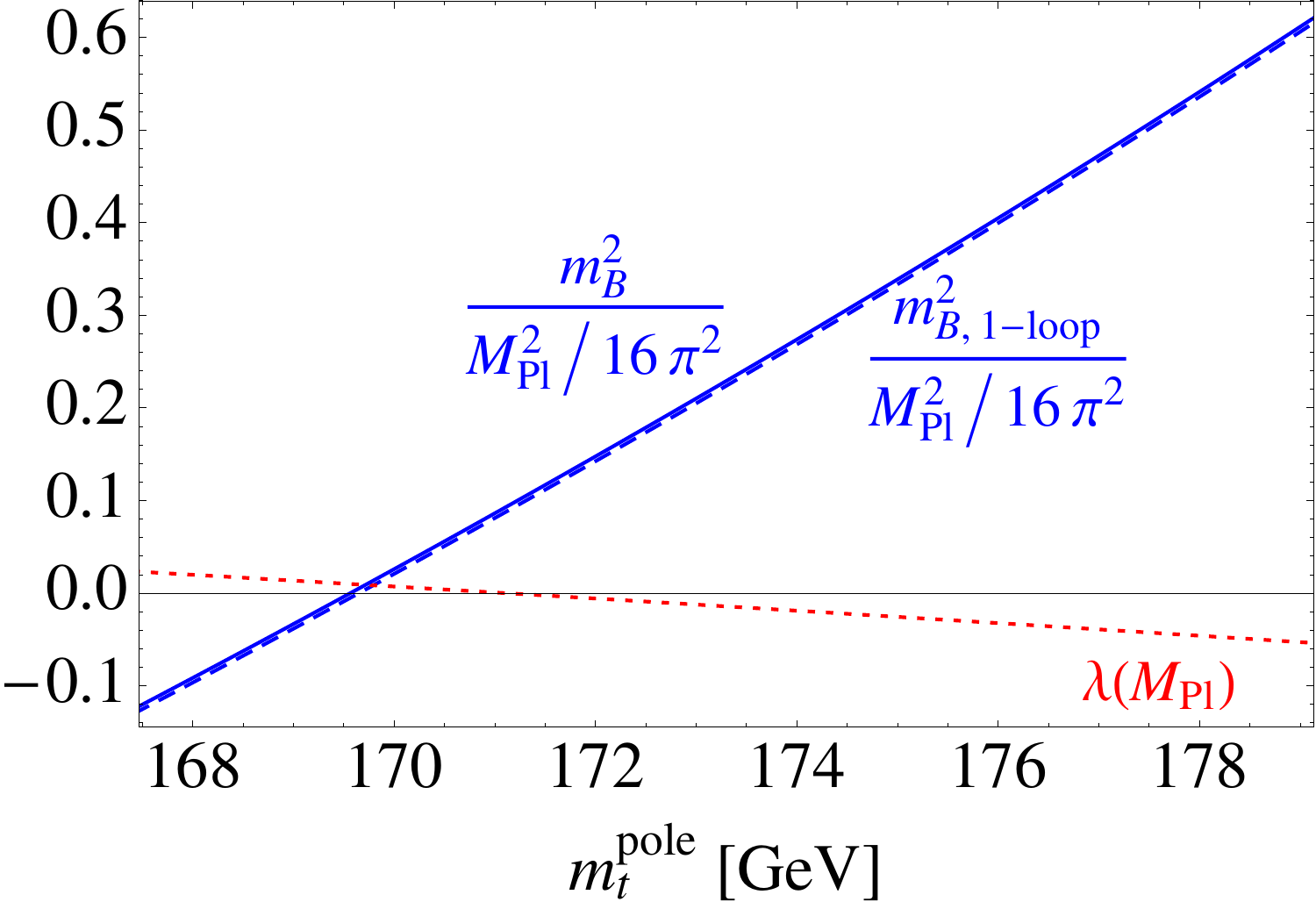}
%  \hfill
%  \includegraphics[width=.35\textwidth]{msq_vs_mt_1.pdf}
%  \hfill
%  \includegraphics[width=.3\textwidth]{msq_vs_mt_2.pdf}
  \hfill
  \mbox{}
	\end{center}
  \caption{
The blue solid (dashed) line corresponds to the one-plus-two-loop (one-loop) bare mass $m_B^2$ ($m_{B,\,\text{1-loop}}^2$) in units of ${M_\text{Pl}^2/16\pi^2}$ for $\Lambda=M_\text{Pl}$. For comparison, we also plot the quartic coupling $\lambda$ at the Planck scale with the red dotted line. The central values $\alpha_s(m_Z)=0.1184$ and $m_H=125.7\GeV$ are used.}
  \label{fig:msq_vs_mt}
%\end{minipage}
\end{figure}
Now we can estimate the bare Higgs mass at the cutoff scale by substituting the $\MSbar$ couplings derived in the previous section to the bare ones in the right-hand sides of Eqs.~\eqref{Veltman 1-loop} and \eqref{bare mass in SM}.

In the left panel of Fig.~\ref{fig:msq}, we plot the dependence of the bare Higgs mass-squared in units of $\Lambda^2/16\pi^2$ on the UV cutoff scale $\Lambda$:
\begin{align}
{m_B^2\over \Lambda^2/16\pi^2}	&=	{m^2_{B,\,\text{1-loop}}\over I_1}+{m^2_{B,\,\text{2-loop}}\over I_2}{I_2\over I_1},
	\label{bare mass}
\end{align}
where we have taken $I_2/I_1=0.005$ as in Eq.~\eqref{two-loop in our scheme}.
In the figure, we can see that the bare mass $m_B^2$ monotonically decreases when one increases $\Lambda$.\footnote{
\addspan{We note again that the bare Higgs mass is defined for each UV cutoff $\Lambda$ and is not a running quantity.}
}

We obtain the UV cutoff scale at which the bare mass $m_B^2$ becomes zero:
\begin{align}
\log_{10}{\left.\Lambda\right|_{m_B^2=0}\over\GeV}
	&=	23.5
		+3.3\left({m_t^\text{pole}-173.3\GeV\over2.8\GeV}\right)
		-0.2\left({m_H-125.7\GeV\over0.6\GeV}\right)\nn
	&\quad
		-0.4\left({\alpha_s(m_Z)-0.1184\over0.0007}\right)
		\pm0.4_\text{th}.
		\label{fine-tuned scale}
\end{align}
In the right panel of Fig.~\ref{fig:msq}, we plot this quantity as a function of the top quark pole mass for the central values of $\alpha_s(m_Z)$ and $m_H$, without referring to the linear approximation~\eqref{fine-tuned scale}.

We show an approximate formula for the bare Higgs mass when the cutoff is at the Planck scale, $\Lambda=M_{\text{Pl}}$:
\begin{align}
m_B^2
	&=	\bigg[
		0.22
		+0.18\left({m_t^\text{pole}-173.3\GeV\over2.8\GeV}\right)
		-0.02\left({\alpha_s(m_Z)-0.1184\over0.0007}\right)\nn
	&\phantom{=	\bigg[}
		-0.01\left({m_H-125.7\GeV\over0.6\GeV}\right)
		\pm0.02_\text{th}\bigg]\frac{M_{\text{Pl}}^2}{16\pi^2}.
		\label{bare mass at Planck scale}
\end{align}
This is one of our main results. 
We verify that the two-loop correction~\eqref{bare mass in SM} can be safely neglected: $m^2_{B,\,\text{2-loop}}\simeq-0.005\,M_\text{Pl}^2/16\pi^2$ within the \addspan{cutoff} scheme~\eqref{two-loop in our scheme}, as advertised before.
%Let us see the value of the bare mass $m_B^2$ at the Planck scale $M_\text{Pl}$. 
In Fig.~\ref{fig:msq_vs_mt}, we plot the bare Higgs mass-squared in units of $M_\text{Pl}^2/16\pi^2$ as a function of $m_t^\text{pole}$ for the central values of $\alpha_s(m_Z)$ and $m_H$, without referring to the linear approximation~\eqref{bare mass at Planck scale}. For comparison, we also plot the quartic coupling $\lambda$ at the Planck scale. 

From Fig.~\ref{fig:msq_vs_mt} we see that the bare Higgs mass becomes zero if $m_t^\text{pole}=169.8\GeV$, while the quartic coupling $\lambda(M_\text{Pl})$ vanishes if $m_t^\text{pole}=171.2\GeV$, when we take the central values for $\alpha_s(m_Z)$ and $m_H$. 
See Refs.~\cite{tuning within QFT} for arguments supporting the vanishing \cred{parameter} at a cutoff scale, see also Ref.~\cite{requested}. 
There is no low energy parameter set within two sigma that makes both the quartic coupling and the bare mass vanish simultaneously at the Planck scale. This might suggest an existence of a small threshold effect from an underlying UV complete theory.

%\begin{figure}[t]
%\begin{minipage}{0.32\hsize}
%  \begin{center}
%   \includegraphics[width=50mm]{0125.pdf}
%  \end{center}
%  \label{0125}
%\end{minipage}
%\begin{minipage}{0.32\hsize}
%\begin{center}
%  \includegraphics[width=50mm]{0126.pdf}
%\end{center}
%  \label{0126}
%\end{minipage}
%\begin{minipage}{0.32\hsize}
%\begin{center}
%  \includegraphics[width=50mm]{0127.pdf}
%\end{center}
%  \label{0127}
%\end{minipage}
%\caption{The contour of the particular UV cutoff scale where the bare mass $m_B^2$ becomes zero. From left to right, $m_H=$ 125, 126, and 127\GeV。The black dot is the central value $(m_t^\text{pole},\alpha_s)=(173.3\GeV,0.1184)$ and the circle is the $1\sigma$ contour from the current experiment. {\bf\color{red} (Need to change from Strumia's!)}}
%\label{msq=0}
%\end{figure}
%
%
%
%
%\begin{figure}[t]
%\begin{minipage}{0.32\hsize}
%  \begin{center}
%   \includegraphics[width=50mm]{125.pdf}
%  \end{center}
%  \label{fig:125}
%\end{minipage}
%\begin{minipage}{0.32\hsize}
%\begin{center}
%  \includegraphics[width=50mm]{126.pdf}
%\end{center}
%  \label{fig:126}
%\end{minipage}
%\begin{minipage}{0.32\hsize}
%\begin{center}
%  \includegraphics[width=50mm]{127.pdf}
%\end{center}
%  \label{fig:127}
%\end{minipage}
%\caption{$M_{\text{Pl}}$での$m_B^2$を$\frac{M_{\text{Pl}}^2}{16 \pi^2}$をunitにして等高線plotしたもの。左から順に$m_H=125,126,127$GeV。黒点は中心値$(m_t^\text{pole},\alpha_s)=(173.3\GeV,0.1184)$で90\%,95\%,99\%の楕円も示した。}
%\label{msqpl}
%\end{figure}

\section{Summary and discussions}
\cred{It is important to fix all the parameters, including the bare Higgs mass, at the UV cutoff scale of the Standard Model in order to explore the Planck scale physics. We note again that in this paper we are not trying to solve the fine-tuning problem but to determine all the bare parameters at the cutoff scale. In addition, we investigate the scale of the vanishing bare mass as a hint of that of the supersymmetry restoration.}

We have presented a \addspan{procedure} where the quadratic divergence of the bare Higgs mass is computed in terms of the bare couplings at a UV cutoff scale $\Lambda$. Using it, we have obtained the bare Higgs mass up to the two-loop order in the SM. This calculation has been made easier by working in the symmetric phase $\left\langle\phi\right\rangle=0$ and in the Landau gauge. We have checked that all the IR divergent terms, which are proportional to $\Lambda^2\ln(\Lambda/\mu_\text{IR})$, cancel out as expected. Approximating the bare couplings at $\Lambda$ by the corresponding $\MSbar$ ones at the same scale, we can examine whether the quadratic divergence in the bare Higgs mass vanishes or not. To get the $\MSbar$ couplings at high scales, we employ the two-loop RGE in the SM. We have found that it is indeed the case if the top quark mass is $m_t^\text{pole}=169.8\GeV$, which is $1.3\,\sigma$ smaller than the current central value.\footnote{
The vanishing of the quadratic divergence does not immediately indicate that the bare Higgs mass is exactly zero. Our result does not exclude logarithmically divergent corrections such as $m_H^2\ln(\Lambda/m_H)$ or finite ones. If the quadratic divergence indeed vanishes exactly for some reason, then such corrections become important. It would be interesting to study them.
}
One might find it intriguing that this value is close to $m_t^\text{pole}=171.2\GeV$, which gives a vanishing quartic coupling at $M_\text{Pl}$.

%There are several arguments~\cite{cosmological constant,Higgs fine-tuning} suggesting that all the couplings, including the bare Higgs mass $m_B^2$, should vanish at a certain UV cutoff scale.
It is a curious fact that the scale of the vanishing bare Higgs mass $m_B^2$ and that for the quartic coupling $\lambda$ are quite close to each other and to the Planck scale. \cred{The fact that the Planck scale appears only from the SM might indicate that the SM is indeed valid up to the Planck scale and is a direct consequence of an underlying physics there. Also, it may imply an almost flat potential near the Planck scale, which opens a possibility that the slow-roll inflation is achieved solely by the Higgs potential.
}

If we take all the central values for $m_t^\text{pole}$, $\alpha_s(m_Z)$, and $m_H$, then the cancellation occurs not at the Planck scale but at a scale around $\Lambda\sim 10^{23}\GeV$. This may hint a new physics around that scale. In this case however, we need to take the graviton effects into account, as discussed in Sec.~\ref{graviton section}.

There can be a different interpretation for the small bare Higgs mass $m_B^2$ left at the Planck scale. It might appear as a threshold correction in string theory. In string theory, the tree-level masses of the particles are quantized by $m_s:=(\alpha')^{-1/2}$, and therefore the Higgs mass is zero at the tree-level. The threshold effect from integrating out the massive stringy excitations is obtained by computing insertions of two Higgs emission vertices with zero external momenta into the world sheet.
The result would become
\begin{equation}
m^2_B
	\sim C\frac{g_s^2}{16\pi^2}m_s^2,
\end{equation}
where $C$ is a model dependent constant. This calculation can be performed for a concrete model such as the orbifold and fermionic constructions in heterotic string. This work will be presented in a separate publication.
%似たような状況でKaplunovskyがgauge couplingへのthreshold correctionを調べている\cite{Kaplunovsky:1987rp}。$m_B^2$は
%この計算を具体的にやって$M_{\text{Pl}}$での$m_B^2$と比べることで, そのmodelでのstring couplingの値を決めることができる。
%これを典型的なstring modelを使ってやることで, 典型的なstring couplingを与えることができるであろう。
%further workとして, このようなstringの1-loop計算に挑戦したい。

We comment on the case where the UV completion of the SM appears as a supersymmetry.
When the supersymmetry is softly broken, there cannot be any quadratic divergence and our study does not apply.
In the case of the split/high-scale supersymmetry~\cite{ArkaniHamed:2004fb,Hall:2009nd} it is possible to perform a parallel analysis to the current one, which will be shown elsewhere.

If we assume the seesaw mechanism, the right-handed neutrinos are introduced above an intermediate scale $M_R$. Our analysis corresponds to the case where $M_R$ is small enough that all the neutrino Dirac-Yukawa couplings are negligible $y_D\lesssim 10^{-1}$. This condition implies $M_R\lesssim 10^{12}\GeV$ for the neutrino mass $m_\nu\sim y_D^2v^2/M_R\sim 0.1\,\text{eV}$. It would be interesting to extend our analysis to include larger Dirac-Yukawa couplings for $M_R\gtrsim 10^{12}\GeV$.

\subsection*{\err{Note added}}
\err{It has been pointed out~\cite{Jones:2013aua} that the formula for the two loop bare Higgs mass~\eqref{bare mass in SM} in the previous version does not agree with the result in Ref.~\cite{Alsarhi:1991ji}. They obtained it from the residue at $d=3$ in the dimensional reduction using the background Feynman gauge, whereas we have computed Eq.~(15) in the Landau gauge in four dimensions. 
The quantity under consideration is an on-shell quantity, namely the two-point function with zero external momentum in the massless theory. Therefore it is a gauge invariant quantity, and hence two results should agree with each other. 
We have re-examined our calculation and found errors that are suggested in Ref.~\cite{Jones:2013aua}. Eq.~\eqref{bare mass in SM} is the corrected version. The error does not influence the consequence, that is, the two loop bare mass is still negligible compared to the one loop one.
}

\subsection*{Acknowledgements}
We greatly appreciate valuable discussions with Satoshi Iso and useful information from him, without which this study would never have been started.
\err{We thank D.~R.~T.~Jones for the kind correspondence.}
This work is in part supported by the Grants-in-Aid for Scientific Research No.~22540277 (HK), No.~23104009, No.~20244028, and No.~23740192 (KO) and for the Global COE program ``The Next Generation of Physics, Spun from Universality and Emergence.''

\appendix
\section*{Appendix}

\section{Cutoff vs $\MSbar$}\label{cutoff vs MSbar}
We have approximated the \addspan{dimensionless} bare coupling constants in the SM by the running ones in the $\MSbar$ scheme at $\Lambda$. 
%With this assumption, we will examine at what cutoff scale the bare Higgs mass becomes zero.
% we will approximate the bare couplings at the UV cutoff scale by the corresponding $\MSbar$ ones at the same scale. 
%Though this is a natural choice, we note that there can be a possible uncertainty. 
The resulting error can be evaluated once the cutoff scheme is explicitly specified. 
%as, say, the momentum cutoff, lattice, Pauli-Villars, and string theory and finite parts of several one-loop diagrams are calculated.
%This uncertainty can be in principle removable if we stick to the SM bare Lagrangian and compute everything by utilizing the bare couplings to sufficient orders, perform resummations of leading logs, and put a renormalization condition in terms of physical quantities.

More concretely, let us first express the $\MSbar$ couplings at a scale $\mu$ in terms of the bare couplings
%\footnote{\addedSecond{In a cutoff regularization, a dimensionless renormalized quantity at a scale $\mu$ is a function of bare couplings $\lambda^i_B$ and of $\log(\mu/\Lambda)$. The coupling $\lambda^i_{\overline{\text{MS}}}(\mu)$ is such a quantity and hence is written as in Eq.~\eqref{higher order explanation}.}} 
defined at the cutoff scale $\Lambda$:
\begin{align}
\lambda^i_{\MSbar}(\mu)
	&=	\lambda^i_B
		+\sum_{jk}c^{ijk}\!\left(\mu/\Lambda\right)\,\lambda_B^j\lambda_B^k
		+O(\lambda_B^3),
				\label{higher order explanation}	\\
c^{ijk}(x)
	&:=	f^{ijk}+b^{ijk}\ln x+O(x^2),
\end{align}
where $b^{ijk}$ is the coefficient in the one-loop beta function and $f^{ijk}$ is the finite part from the one-loop diagrams. $\left\{\lambda^i_{\MSbar}\right\}_{i=1,\dots,5}$ ($\left\{\lambda^i_B\right\}_{i=1,\dots,5}$) stands for the $\MSbar$ (bare) couplings of the SM: $\left\{g_Y^2, g_2^2, g_3^2, y_t^2, \lambda\right\}$ ($\left\{g_{YB}^2, g_{2B}^2, g_{3B}^2, y_{tB}^2, \lambda_B\right\}$).

In our case, the two-loop corrections in the RGE at high scales are small compared to the one-loop order, which indicates that the two-loop terms $O(\lambda_B^3)$ in Eq.~\eqref{higher order explanation} are negligible as we can take $\mu$ that satisfies both
\begin{align}
\mu
	&\ll	 \Lambda,	&
\left|{\lambda^i_{\MSbar}\over16\pi^2}\ln(\mu/\Lambda)\right| 
	&\ll	1,
\end{align}
simultaneously. Thus we have 
\begin{align}
\lambda^i_{\MSbar}(\mu)
	&=	\lambda^i_B
		+\sum_{jk}\left(f^{ijk}+b^{ijk}\ln{\mu\over\Lambda}\right)\,\lambda_B^j\lambda_B^k.
			\label{MSbar approx}
\end{align}
On the other hand, from the RGE, we get
\begin{align}
\lambda^i_{\MSbar}(\Lambda)
	&=	\lambda^i_{\MSbar}(\mu)
		+\sum_{jk}b^{ijk}
		\lambda_{\MSbar}^j(\mu)\,
		\lambda_{\MSbar}^k(\mu)
		\ln{\Lambda\over\mu}.
			\label{MSbar Lambda and mu}
\end{align}
From Eqs.~\eqref{MSbar approx} and \eqref{MSbar Lambda and mu}, we obtain
\begin{align}
\lambda^i_{\MSbar}(\Lambda)
	&=	\lambda^i_B+\sum_{jk}f^{ijk}\lambda^j_B\lambda^k_B,
\end{align}
which gives the relation between the bare and the $\MSbar$ couplings at the same scale.

With the above correction, the formula for the bare Higgs mass is modified by
\begin{align}
\Delta m_B^2
	&=	-\sum_{ijk}a^if^{ijk}\lambda^j_{\MSbar}(\Lambda)\,\lambda^k_{\MSbar}(\Lambda),
\end{align}
where $a^i$ are the coefficients in the one-loop bare Higgs mass $m_B^2=\sum_ia^i\lambda^i_B$ in Eq.~\eqref{Veltman 1-loop}, and are proportional to $I_1$.
The scale at which the bare Higgs mass vanishes $\left.\Lambda\right|_{m_B^2=0}$ is changed to $\left.\Lambda\right|_{m_B^2=0}e^{\delta t}$, where
\begin{align}
\delta t
	&=	{\sum_{ijk}a^if^{ijk}\lambda^j_{\MSbar}(\Lambda)\,\lambda^k_{\MSbar}(\Lambda)\over
		\sum_{ijk}a^ib^{ijk}\lambda^j_{\MSbar}(\Lambda)\,\lambda^k_{\MSbar}(\Lambda)}.
\end{align}
Generically $f^{ijk}$ are of the same order as $b^{ijk}$ and hence the correction due to the replacement of the bare couplings by the $\MSbar$ ones, $\Delta m_B^2$, is as small as the two-loop corrections. Since $\delta t$ is of order unity, the ambiguity for the scale $\left.\Lambda\right|_{m_B^2=0}$ would be at most $e^{\delta t}\lesssim 10$.


\begin{thebibliography}{99}

%\cite{:2012gk}
\bibitem{:2012gk}
  G.~Aad {\it et al.}  [ATLAS Collaboration],
  {\it ``Observation of a new particle in the search for the Standard Model Higgs boson with the ATLAS detector at the LHC,''}  Phys.\ Lett.\ B {\bf 716} (2012) 1  [arXiv:1207.7214 [hep-ex]].  %%CITATION = ARXIV:1207.7214;%%

%\cite{:2012gu}
\bibitem{:2012gu}
  S.~Chatrchyan {\it et al.}  [CMS Collaboration],
  {\it ``Observation of a new boson at a mass of 125 GeV with the CMS experiment at the LHC,''}  Phys.\ Lett.\ B {\bf 716} (2012) 30  [arXiv:1207.7235 [hep-ex]].  %%CITATION = ARXIV:1207.7235;%%

\bibitem{Beringer:1900zz}
  J.~Beringer {\it et al.}  [Particle Data Group Collaboration],
  {\it ``Review of Particle Physics (RPP),''}
  Phys.\ Rev.\ D {\bf 86} (2012) 010001.
  %%CITATION = PHRVA,D86,010001;%%

\bibitem{SM stability}
%\bibitem{Holthausen:2011aa}
  M.~Holthausen, K.~S.~Lim and M.~Lindner,
  {\it ``Planck scale Boundary Conditions and the Higgs Mass,''}
JHEP {\bf 1202} (2012) 037  [arXiv:1112.2415 [hep-ph]];
%%CITATION = ARXIV:1112.2415;%%
	\\
%\bibitem{Bezrukov:2012sa}
  F.~Bezrukov, M.~Y.~Kalmykov, B.~A.~Kniehl and M.~Shaposhnikov,
  {\it ``Higgs boson mass and new physics,''}
  arXiv:1205.2893 [hep-ph];
  %%CITATION = ARXIV:1205.2893;%%
  \\
%\bibitem{Masina:2012tz}
  I.~Masina,
  {\it ``The Higgs boson and Top quark masses as tests of Electroweak Vacuum Stability,''}
  arXiv:1209.0393 [hep-ph].
  %%CITATION = ARXIV:1209.0393;%%

\bibitem{Degrassi:2012ry}
  G.~Degrassi, S.~Di Vita, J.~Elias-Miro, J.~R.~Espinosa, G.~F.~Giudice, G.~Isidori and A.~Strumia,
  {\it ``Higgs mass and vacuum stability in the Standard Model at NNLO,''}
  JHEP {\bf 1208} (2012) 098
  [arXiv:1205.6497 [hep-ph]].
  %%CITATION = ARXIV:1205.6497;%%

\bibitem{Alekhin:2012py}
  S.~Alekhin, A.~Djouadi and S.~Moch,
  {\it ``The top quark and Higgs boson masses and the stability of the electroweak vacuum,''}
  Phys.\ Lett.\ B {\bf 716} (2012) 214
  [arXiv:1207.0980 [hep-ph]].
  %%CITATION = ARXIV:1207.0980;%%






\bibitem{Martin:1997ns}
  S.~P.~Martin,
  {\it ``A Supersymmetry primer,''}
  In *Kane, G.L. (ed.): Perspectives on supersymmetry II* 1-153
  [hep-ph/9709356].
  %%CITATION = HEP-PH/9709356;%%

\bibitem{Baer:2012uy}
  H.~Baer, V.~Barger, P.~Huang and X.~Tata,
  {\it ``Natural Supersymmetry: LHC, dark matter and ILC searches,''}
  JHEP {\bf 1205} (2012) 109
  [arXiv:1203.5539 [hep-ph]].
  %%CITATION = ARXIV:1203.5539;%%

\bibitem{nosusy}
 The ATLAS Collaboration,
 {\it ``Search for new phenomena using large jet multiplicities and missing transverse momentum with ATLAS in 5.8 fb$^{-1}$ of $\sqrt{s}=8$\,TeV proton-proton collisions,''} ATLAS-CONF-2012-103; \\
 {\it ``Search for supersymmetry at $\sqrt{s}=8$\,TeV in final states with jets, missing transverse momentum and one isolated lepton,''}
 ATLAS-CONF-2012-104; \\
 {\it ``Search for Supersymmetry in final states with two same-sign leptons, jets and missing transverse momentum with the ATLAS detector in pp collisions at $\sqrt{s}=8$\,TeV,''}
 ATLAS-CONF-2012-105; \\
 {\it ``Search for squarks and gluinos with the ATLAS detector using final states with jets and missing transverse momentum at $\sqrt{s} = 8$\,TeV,''}
 ATLAS-CONF-2012-109;\\
 The CMS Collaboration, 
 {\it ``Search for supersymmetery in final states with missing transverse momentum and 0, 1, 2, or $\ge$3 b jets in 8 TeV pp collisions''}
 CMS-PAS-SUS-12-016; \\
 {\it ``Search for New Physics in Events with a Z Boson and Missing Transverse Energy,''}
 CMS-PAS-SUS-12-017; \\
 {\it ``Search for supersymmetry in events with photons and missing energy,''}
 CMS-PAS-SUS-12-018.


\bibitem{Weinberg:1987dv}
  S.~Weinberg,
  {\it ``Anthropic Bound on the Cosmological Constant,''}
  Phys.\ Rev.\ Lett.\  {\bf 59} (1987) 2607.
  %%CITATION = PRLTA,59,2607;%%

\bibitem{Hall:2009nd}
  L.~J.~Hall and Y.~Nomura,
  {\it ``A Finely-Predicted Higgs Boson Mass from A Finely-Tuned Weak Scale,''}
  JHEP {\bf 1003} (2010) 076
  [arXiv:0910.2235 [hep-ph]].
  %%CITATION = ARXIV:0910.2235;%%


\bibitem{multiverse}
%\bibitem{Linde:1988ws}
  A.~D.~Linde,
  {\it ``The Universe Multiplication And The Cosmological Constant Problem,''}
  Phys.\ Lett.\ B {\bf 200} (1988) 272;
  %%CITATION = PHLTA,B200,272;%%
  \\
%\bibitem{Coleman:1988tj}
  S.~R.~Coleman,
  {\it ``Why There Is Nothing Rather Than Something: A Theory of the Cosmological Constant,''}
  Nucl.\ Phys.\ B {\bf 310} (1988) 643;
  %%CITATION = NUPHA,B310,643;%%
  \\
%\bibitem{Weinberg:1988cp}
  S.~Weinberg,
  {\it ``The Cosmological Constant Problem,''}
  Rev.\ Mod.\ Phys.\  {\bf 61} (1989) 1;
  %%CITATION = RMPHA,61,1;%%
  \\
  H.~Kawai and T.~Okada,
  {\it ``Solving the Naturalness Problem by Baby Universes in the Lorentzian Multiverse,''}
  Prog.\ Theor.\ Phys.\  {\bf 127} (2012) 689
  [arXiv:1110.2303 [hep-th]].
  %%CITATION = ARXIV:1110.2303;%%


\bibitem{tuning within QFT}
%\bibitem{Froggatt:1995rt}
  C.~D.~Froggatt and H.~B.~Nielsen,
  {\it ``Standard model criticality prediction: Top mass 173 $\pm$ 5-GeV and Higgs mass 135 +- 9-GeV,''}
  Phys.\ Lett.\ B {\bf 368} (1996) 96
  [hep-ph/9511371];\\
  %%CITATION = HEP-PH/9511371;%%
%\bibitem{Borstnik:1999wu}
  B.~Stech in
  {\it ``Proceedings to the Workshop at Bled, Slovenia, 29 June - 9 July 1998: What comes beyond the Standard Model,''}
  Ljubljana, Slovenia: DMFA (1999) 73 p
  [hep-ph/9905357];\\
  %%CITATION = HEP-PH/9905357;%%
%\bibitem{Meissner:2006zh}
  K.~A.~Meissner and H.~Nicolai,
  {\it ``Conformal Symmetry and the Standard Model,''}
  Phys.\ Lett.\ B {\bf 648} (2007) 312
  [hep-th/0612165];
  %%CITATION = HEP-TH/0612165;%%
%\bibitem{Meissner:2007xv}
  %K.~A.~Meissner and H.~Nicolai,
  {\it ``Effective action, conformal anomaly and the issue of quadratic divergences,''}
  Phys.\ Lett.\ B {\bf 660} (2008) 260
  [arXiv:0710.2840 [hep-th]];\\
  %%CITATION = ARXIV:0710.2840;%%
%\bibitem{Iso:2009ss}
  S.~Iso, N.~Okada and Y.~Orikasa,
  {\it ``Classically conformal $B-L$ extended Standard Model,''}
  Phys.\ Lett.\ B {\bf 676} (2009) 81
  [arXiv:0902.4050 [hep-ph]];
  %%CITATION = ARXIV:0902.4050;%%
%\bibitem{Iso:2009nw}
%  S.~Iso, N.~Okada and Y.~Orikasa,
  {\it ``The minimal $B-L$ model naturally realized at TeV scale,''}
  Phys.\ Rev.\ D {\bf 80} (2009) 115007
  [arXiv:0909.0128 [hep-ph]];\\
  %%CITATION = ARXIV:0909.0128;%%
%\bibitem{Kawai:2011qb}
%\bibitem{Shaposhnikov:2009pv}
  M.~Shaposhnikov and C.~Wetterich,
  {\it ``Asymptotic safety of gravity and the Higgs boson mass,''}
  Phys.\ Lett.\ B {\bf 683} (2010) 196
  [arXiv:0912.0208 [hep-th]].
  %%CITATION = ARXIV:0912.0208;%%





%\bibitem{Weinberg:1951ss}
%  \addspan{
%  S.~Weinberg,
%  {\it ``New approach to the renormalization group,''}
%  Phys.\ Rev.\ D {\bf 8} (1973) 3497.
%  %%CITATION = PHRVA,D8,3497;%%
%  }





%\cite{Veltman:1980mj}
\bibitem{Veltman:1980mj}
  M.~J.~G.~Veltman,
  {\it ``The Infrared - Ultraviolet Connection,''}
  Acta Phys.\ Polon.\ B {\bf 12} (1981) 437.
  %%CITATION = APPOA,B12,437;%%





\bibitem{others}
  M.~B.~Einhorn and D.~R.~T.~Jones,
  {\it ``The Effective potential and quadratic divergences,''}
  Phys.\ Rev.\ D {\bf 46} (1992) 5206; \\
  %%CITATION = PHRVA,D46,5206;%%
  C.~F.~Kolda and H.~Murayama,
  {\it ``The Higgs mass and new physics scales in the minimal standard model,''}
  JHEP {\bf 0007} (2000) 035
  [hep-ph/0003170]; \\
  %%CITATION = HEP-PH/0003170;%%
  J.~A.~Casas, J.~R.~Espinosa and I.~Hidalgo,
  {\it ``Implications for new physics from fine-tuning arguments. 1. Application to SUSY and seesaw cases,''}
  JHEP {\bf 0411} (2004) 057
  [hep-ph/0410298].


\bibitem{Drozd:2012is}
  %%CITATION = HEP-PH/0410298;%%
  A.~Drozd,
  {\it ``RGE and the Fine-Tuning Problem,}
  arXiv:1202.0195 [hep-ph].
  %%CITATION = ARXIV:1202.0195;%%

\bibitem{Kawai:1986va}
  H.~Kawai, D.~C.~Lewellen and S.~H.~H.~Tye,
  {\it ``Construction of Four-Dimensional Fermionic String Models,''}
  Phys.\ Rev.\ Lett.\  {\bf 57} (1986) 1832
   [Erratum-ibid.\  {\bf 58} (1987) 429].
  %%CITATION = PRLTA,57,1832;%%


%\bibitem{ours}
%  Y.~Hamada, H.~Kawai and K.~Oda,
%  to appear.

\bibitem{ours}
\err{
Y.~Hamada, H.~Kawai and K.~Oda,
{\it ``Minimal Higgs Inflation,''}
arXiv:1308.6651 [hep-ph].}
%%CITATION = ARXIV:1308.6651;%%

\bibitem{Fujikawa:2011zf}
\addedSecond{
  K.~Fujikawa,
  {\it ``Remark on the subtractive renormalization of quadratically divergent scalar mass,''}
  Phys.\ Rev.\ D {\bf 83} (2011) 105012
  [arXiv:1104.3396 [hep-th]].
  %%CITATION = ARXIV:1104.3396;%%
}
  



\bibitem{Jones:2013aua}
\err{
D.~R.~T.~Jones,
{\it ``The Quadratic Divergence in the Higgs Mass Revisited,''}
Phys.\ Rev.\ D {\bf 88} (2013) 098301
[arXiv:1309.7335 [hep-ph]].
%%CITATION = ARXIV:1309.7335;
}


\bibitem{Alsarhi:1991ji}
\err{M.~S.~Al-sarhi, I.~Jack and D.~R.~T.~Jones,
{\it ``Quadratic Divergences in Gauge Theories,''}
Z.\ Phys.\ C {\bf 55} (1992) 283.}
%%CITATION = ZEPYA,C55,283;%%



%\cite{Arason:1991ic}
\bibitem{Arason:1991ic}
  H.~Arason, D.~J.~Castano, B.~Keszthelyi, S.~Mikaelian, E.~J.~Piard, P.~Ramond and B.~D.~Wright,
  {\it ``Renormalization group study of the standard model and its extensions. 1. The Standard model,''}
  Phys.\ Rev.\ D {\bf 46} (1992) 3945.
  %%CITATION = PHRVA,D46,3945;%%

\bibitem{Ford:1992pn}
  C.~Ford, I.~Jack and D.~R.~T.~Jones,
  {\it ``The Standard model effective potential at two-loops,''}
  Nucl.\ Phys.\ B {\bf 387} (1992) 373
   [Erratum-ibid.\ B {\bf 504} (1997) 551]
  [hep-ph/0111190].
  %%CITATION = HEP-PH/0111190;%%





\bibitem{requested}
%\cite{Foot:2007as}
%\bibitem{Foot:2007as}
R.~Foot, A.~Kobakhidze and R.~R.~Volkas,
{\it ``Electroweak Higgs as a pseudo-Goldstone boson of broken scale invariance,''}
Phys.\ Lett.\ B {\bf 655}, 156 (2007)
[arXiv:0704.1165 [hep-ph]];\\
 %%CITATION = ARXIV:0704.1165;%%
%\cite{Foot:2007iy}
%\bibitem{Foot:2007iy}
R.~Foot, A.~Kobakhidze, K.~L.~McDonald and R.~R.~Volkas,
{\it ``A Solution to the hierarchy problem from an almost decoupled hidden sector within a classically scale invariant theory,''}
Phys.\ Rev.\ D {\bf 77}, 035006 (2008)
[arXiv:0709.2750 [hep-ph]].
%%CITATION = ARXIV:0709.2750;%%

















%%\cite{Melnikov:2000qh}
%\bibitem{Melnikov:2000qh}
%  K.~Melnikov and T.~v.~Ritbergen,
%  %``The Three loop relation between the MS-bar and the pole quark masses,''
%  Phys.\ Lett.\ B {\bf 482} (2000) 99
%  [hep-ph/9912391].
%  %%CITATION = HEP-PH/9912391;%%
%  K.~G.~Chetyrkin and M.~Steinhauser,
%  %``The Relation between the MS-bar and the on-shell quark mass at order alpha(s)**3,''
%  Nucl.\ Phys.\ B {\bf 573} (2000) 617
%  [hep-ph/9911434].
%  %%CITATION = HEP-PH/9911434;%% 

%%\cite{Gray:1990yh}
%\bibitem{Gray:1990yh}
%  N.~Gray, D.~J.~Broadhurst, W.~Grafe and K.~Schilcher,
%  %``THREE LOOP RELATION OF QUARK (modified) MS AND POLE MASSES,''
%  Z.\ Phys.\ C {\bf 48} (1990) 673.
%  %%CITATION = ZEPYA,C48,673;%%

%%\cite{Hempfling:1994ar}
%\bibitem{Hempfling:1994ar}
%  R.~Hempfling and B.~A.~Kniehl,
%  %``On the relation between the fermion pole mass and MS Yukawa coupling in the standard model,''
%  Phys.\ Rev.\ D {\bf 51} (1995) 1386
%  [hep-ph/9408313].
%  %%CITATION = HEP-PH/9408313;%%

%  %\cite{Jegerlehner:2003py}
%\bibitem{Jegerlehner:2003py}
%  F.~Jegerlehner and M.~Y.~.Kalmykov,
%  %``O(alpha alpha(s)) correction to the pole mass of the t quark within the standard model,''
%  Nucl.\ Phys.\ B {\bf 676} (2004) 365
%  [hep-ph/0308216].
%  %%CITATION = HEP-PH/0308216;%%

%  %\cite{Bethke:2009jm}
%\bibitem{Bethke:2009jm}
%  S.~Bethke,
%  %``The 2009 World Average of alpha(s),''
%  Eur.\ Phys.\ J.\ C {\bf 64} (2009) 689
%  [arXiv:0908.1135 [hep-ph]].
%  %%CITATION = ARXIV:0908.1135;%%


%%\cite{Lancaster:2011wr}
%\bibitem{Lancaster:2011wr}
%  [Tevatron Electroweak Working Group and CDF and D0 Collaborations],
%  %``Combination of CDF and D0 results on the mass of the top quark using up to 5.8~fb-1 of data,'' 
%arXiv:1107.5255 [hep-ex].  %%CITATION = ARXIV:1107.5255;%%
%
%
%\bibitem{CMS}
%CMS collaboration,
%CMS-PAS-TOP-11-015
%
%
%\bibitem{ATLAS}
%ATLAS collaboration,
%Top quark mass measurements at the ATLAS experiment
%



%\cite{Kaplunovsky:1987rp}
%\bibitem{Kaplunovsky:1987rp}
%  V.~S.~Kaplunovsky,
%  %``One Loop Threshold Effects in String Unification,'' 
%Nucl.\ Phys.\ B {\bf 307} (1988) 145   [Erratum-ibid.\ B {\bf 382} (1992) 436]  [hep-th/9205068].
%%%CITATION = HEP-TH/9205068;%%


\bibitem{ArkaniHamed:2004fb}
  N.~Arkani-Hamed and S.~Dimopoulos,
  {\it ``Supersymmetric unification without low energy supersymmetry and signatures for fine-tuning at the LHC,''}
  JHEP {\bf 0506} (2005) 073
  [hep-th/0405159].
  %%CITATION = HEP-TH/0405159;%%
  

\end{thebibliography}
\end{document}